\documentclass[aps,prl,superscriptaddress,twocolumn,floatfix,preprintnumbers]{revtex4}
 
\usepackage{amsmath}
\usepackage{graphicx,tabularx,epsfig}
 
\def\beq{\begin{equation}}
\def\eeq{\end{equation}}

\def\bea{\arraycolsep .1em \begin{eqnarray}}
\def\eea{\end{eqnarray}}
\def\Tr{{\rm Tr}}
\def\step{\\[-1.5ex]}
 
\def\eq#1{(\ref{#1})}

\def\s0#1#2{\mbox{\small{$ \frac{#1}{#2} $}}}
\def\0#1#2{\frac{#1}{#2}}

\def\eg{{\it e.g.\;}}

\renewcommand{\mp}{M_D}

\newcommand{\tev}{~{\ensuremath\rm TeV}}

\newcommand{\ifb}{~{\ensuremath\rm fb}^{-1}}
 
\begin{document}

\title{Signatures of gravitational fixed points at the LHC}

\preprint{CERN-TH-PH-2007-127}
 
\author{Daniel F. Litim}
\affiliation{Department of Physics and Astronomy, University of Sussex, 
  Brighton, BN1 9QH, U.K.\\ 
  and Theory Group, Physics Division, CERN, 1211 Geneva 23, Switzerland.}  
\author{Tilman Plehn}
\affiliation{Heisenberg Fellow, SUPA, School of Physics, University of
  Edinburgh, Scotland.}
 
\begin{abstract}
  We study quantum-gravitational signatures at the Large Hadron Collider (LHC)
  in the context of theories with extra spatial dimensions and a low
  fundamental Planck scale in the TeV range. Implications of a gravitational
  fixed point at high energies are worked out using Wilson's renormalisation
  group. We find that relevant cross-sections involving virtual gravitons
  become finite.  Based on gravitational lepton pair production we conclude
  that the LHC is sensitive to a fundamental Planck scale of up to 
  $6$ TeV.
\end{abstract}

\maketitle
 

{\it Introduction.---} The fascinating idea that the fundamental scale of
gravity, the Planck scale, is connected to the scale of electroweak symmetry
breaking -- within a higher dimensional setting -- has stimulated extensive
model building and phenomenology \cite{add}.
Central to these scenarios is that gravity propagates in the higher
dimensional bulk, while Standard Model (SM) particles are often confined to
the four-dimensional brane.  Most interestingly, the hypothesis of a
fundamental Planck scale in the TeV energy range can be tested at high-energy
colliders.  In this way, the LHC becomes sensitive to the dynamics of gravity,
and could possibly become the first experiment able to establish evidence for
the quantisation of gravity.

The phenomenology of quantum gravity at hadron colliders
\cite{grw,real_kk,virtual_kk} is based on direct graviton production like $pp
\to {\rm jet} + G$, leading to missing energy signatures; virtual graviton
exchange in Standard Model reference processes like $pp \to \ell^+\ell^-$,
leading to deviations in critical distributions; and the production and decay
of mini-black holes.  The first two processes are standard search channels for
physics beyond the SM.  In the context of low-scale quantum gravity, they have
been studied within effective field theory \cite{grw}, which allows for a
well controlled description as long as the relevant momentum scales are
sufficiently below an ultraviolet (UV) cutoff of the order of the fundamental
Planck scale $M_D$.  Near the Planck scale and above, an understanding of
gravitational interactions requires an explicit quantum theory for gravity.
It has been suggested that a local quantum theory of gravity in terms of the
metric field may very well exist on a non-perturbative level
\cite{Weinberg79}, despite its perturbative non-renormalisability.  This
asymptotic safety scenario requires the existence of a non-trivial UV fixed
point for quantum gravity under the renormalisation group (RG). In four
dimensions, growing evidence for an UV fixed point has arisen recently, based
on RG studies
\cite{Reuter:1996cp,Souma:1999at,Lauscher:2001rz,Percacci:2003jz,Litim:2003vp}
(see \cite{Niedermaier:2006ns} for reviews) and lattice simulations
\cite{lattice}.  This picture has also been extended to higher dimensions
\cite{Litim:2003vp,Fischer:2006fz}.  Hence, asymptotically safe gravity
provides an excellent starting point to access the quantum gravitational
domain.

In this Letter, we study implications of gravitational fixed points on lepton
pair production $pp\to \ell^+\ell^-$ through graviton exchange in scenarios
with large extra dimensions. The main new ingredient is the non-trivial RG
running of the gravitational sector in higher dimensions
\cite{Litim:2003vp,Fischer:2006fz}. We first discuss basic implications of a
gravitational fixed point.  We then show that fixed point scaling leads to a
finite virtual graviton amplitude.  We compute the effective cross sections
and the $5\sigma$ discovery reach for the fundamental Planck scale at LHC
energies.

{\it Gravitational fixed point.---} We begin with a discussion of structural
implications of gravitational fixed points and consider a Callan-Symanzik type
equation for the gravitational coupling $G$ with respect to a momentum scale
$\mu$ in $D$ dimensions \cite{Litim:2003vp,Niedermaier:2006ns}. In terms of
the running dimensionless gravitational coupling $g(\mu)=G(\mu)\mu^{D-2}\equiv
G_0 Z_G^{-1}(\mu)\mu^{D-2}$, it reads
\begin{equation}\label{dg}
\beta_g\equiv
\frac{{\rm d}\, g(\mu)}{{\rm d}\ln \mu} = (D-2+\eta)g(\mu)\,.
\end{equation}
Here $\eta=-\mu\partial_\mu \ln Z_G$ denotes the anomalous dimension of
the graviton. We assume that the fundamental action is local in the metric
field.  The wave function factor is normalised as $Z_G(\mu_0)=1$ at some
reference scale $\mu_0$ with $G(\mu_0)$ given by Newton's constant $G_0$. In
general, the anomalous dimension depends on all couplings of the theory. Due
to its structure, \eq{dg} can achieve two types of fixed points. At small
coupling, the anomalous dimension vanishes and $g=0$ corresponds to the
non-interacting ($i.e.$~Gaussian) fixed point of \eq{dg}. This fixed point
dominates the deep infrared (IR) region of gravity where $\mu\to 0$. In turn,
an interacting fixed point $g_*$ may be achieved if the anomalous dimension of
the graviton becomes non-perturbatively large,
\begin{equation}\label{eta}
\eta_* = 2-D.
\end{equation} 
Hence, a non-trivial fixed point of quantum gravity in $D>2$ implies a
negative integer value for the graviton anomalous dimension, precisely
counter-balancing the canonical dimension of $G$.  This behaviour implies that
the gravitational coupling constant scales as $G(\mu)\to g_*/\mu^{D-2}$ in the
vicinity of a non-trivial fixed point. In the UV limit where $\mu\to \infty$,
the gravitational coupling $G(\mu)$ becomes arbitrarily weak.
\step

{\it Wilson's renormalisation group.---} Now we turn to RG flows for gravity
\cite{Reuter:1996cp} and consider the effective action $\Gamma_k$
\begin{equation}\label{EHk}
 \Gamma_k=
\0{1}{16\pi G_k}\int d^Dx \sqrt{g}\left[-R(g)+\cdots\right]
\end{equation}
where $k$ denotes the Wilsonian RG scale replacing the scale $\mu$ introduced
in \eq{dg}, and $R(g)$ denotes the Ricci scalar. The dots in \eq{EHk} stand
for the cosmological constant, higher dimensional operators in the metric
field, gravity-matter interactions, a classical gauge fixing and ghost terms.
In Wilson's approach, all couplings in \eq{EHk} become running couplings as
functions of the momentum scale $k$.  For $k\ll M_D$, the gravitational sector
is well-approximated by the Einstein-Hilbert action with $G_k\approx G_0$, and
similarily for the gravity-matter couplings. The corresponding operators scale
canonically.  At $k\approx M_D$ and above, the non-trivial RG running of
gravitational couplings becomes important. The Wilsonian RG flow for an action
\eq{EHk} is given by an exact differential equation
\cite{Wetterich:1992yh,Reuter:1996cp}
\beq\label{ERG} 
\partial_t \Gamma_k=
\frac{1}{2} \Tr \left({\Gamma_k^{(2)}+R_k}\right)^{-1}\partial_t R_k
\eeq 
and $t=\ln k$. The trace stands for a momentum integration and a sum over
indices and fields, and $R_k(q^2)$ denotes an appropriate infrared cutoff
function at momentum scale $q^2\approx k^2$ \cite{Litim:2000ci}.
Diffeomorphism invariance under local coordinate transformations is controlled
by modified Ward identities \cite{Reuter:1996cp}, similar to those known for
non-Abelian gauge theories \cite{Freire:2000bq}.  \step

\begin{center}
\begin{figure}
  \unitlength0.001\hsize
\begin{picture}(1000,360)
\put(850,280){{\Large $|\eta|$}}
\put(850,150){{$n=1$}}
\put(320,310){{$n=7$}}
\epsfig{file=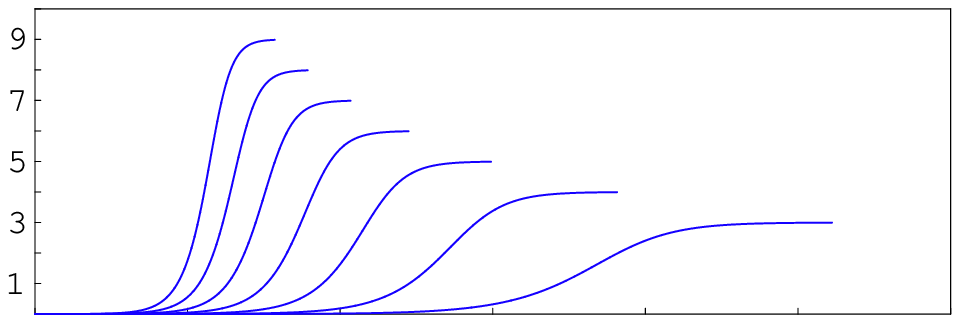,width=\hsize}
\end{picture}
\begin{picture}(1000,310)
\put(850,250){{\large $\displaystyle \frac{G_k}{G_0}$}}
\put(480,-30){{\large $\ln k$}}
\epsfig{file=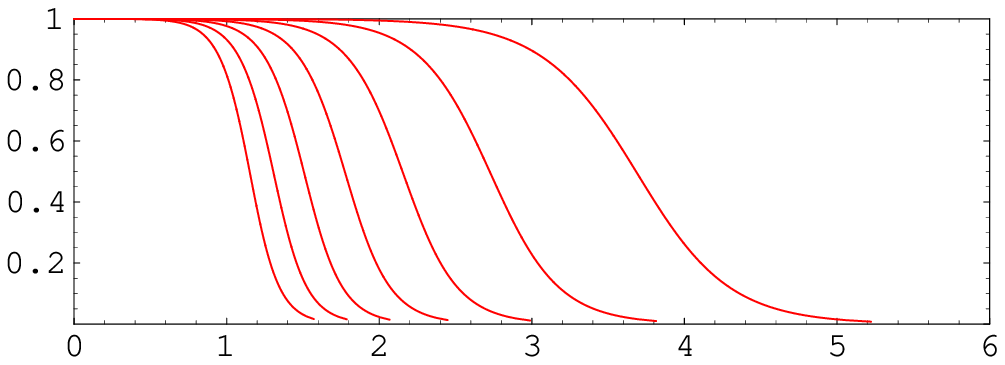,width=\hsize}
\end{picture}
\vskip.2cm
 \caption{\label{Running}
   Running anomalous dimensions (upper panel) and gravitational couplings
   (lower panel) from \eq{betag0} and \eq{g0-explicit} for $D=4+n$ dimensions
   with $n=1,\cdots,7$ (from right to left). The cross-over takes place at the
   respective Planck scale.}\vskip-.3cm
\end{figure}
\end{center}

{\it Running couplings.---} To illustrate the main RG effects we approximate
\eq{EHk} by the Ricci scalar and discuss the running of $g_k$, following
\cite{Litim:2003vp}. The inclusion of a cosmological constant modifies the
approach towards fixed point scaling \cite{Fischer:2006fz} without altering
the central pattern relevant for the discussion below.  Using \eq{EHk} and
\eq{ERG}, we find \cite{Litim:2003vp}
\beq\label{betag0}
\beta_g=\0{(1-4Dg)(D-2)g}{1-(2D-4)g}\,
\quad
\eta=\0{2(D-2)(D+2)\,g}{2(D-2)\,g-1}
\eeq
where $g$ has been rescaled by a numerical factor. We observe a Gaussian fixed
point, and a non-Gaussian one at $g_*=1/(4D)$.  Integrating the flow
\eq{betag0} gives
\beq\label{g0-explicit}
\frac{1}{D-2}\ln\left|\0{g_k}{g_0}\right|
-\0{1}{\theta_{\rm NG}}\ln\left|\0{g_*-\,g_k}{g_*-\,g_0}\right|
=
\ln\0{k}{k_0}\,
\eeq
with initial condition $g_0$ at $k=k_0$, and $\theta_{\rm NG}=2D\ 
\0{D-2}{D+2}$.  The result \eq{g0-explicit} holds for generic momentum cutoff,
only $g_*$ and the scaling exponent $\theta_{\rm NG}$ depend slightly on the
cutoff choice \cite{Litim:2003vp,Fischer:2006fz}.  Inserting the running
coupling \eq{g0-explicit} into \eq{betag0} shows that the anomalous dimension
displays a smooth cross-over between the IR domain $k\ll k_{\rm tr}$ where
$\eta \approx 0$ and the UV domain $k\gg k_{\rm tr}$ where $\eta
\approx2-D$ (see Fig.~\ref{Running}), and $k$ appropriately normalised with
$k_{\rm tr}$ of the order of the Planck scale.  We note that the cross-over
regime becomes narrower with increasing
dimension.  \\[-8ex]

\begin{figure*}[t]
\includegraphics[width=.95\textwidth]{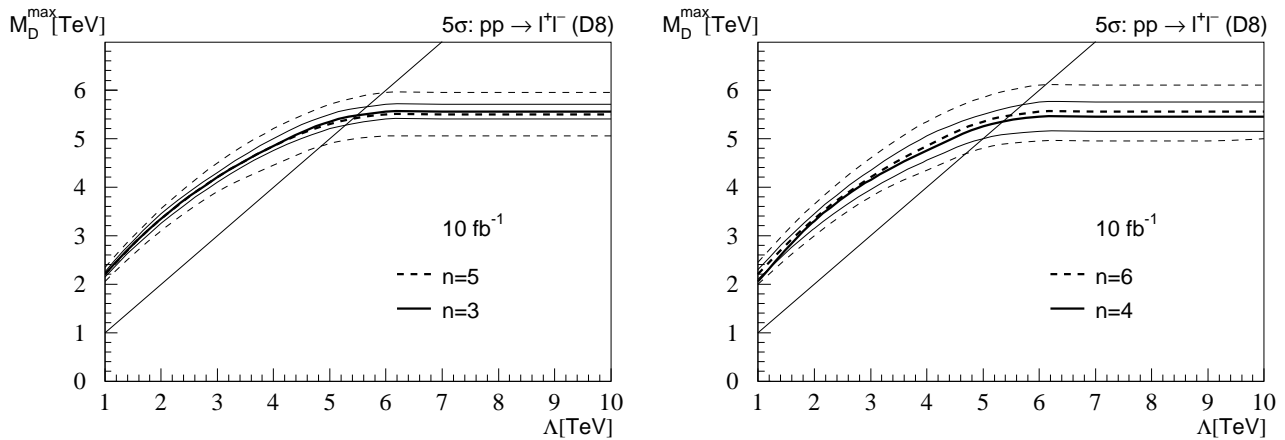}\vskip-.4cm
\caption{The $5\sigma$ discovery contours in $M_D$ 
  at the LHC, shown as a function of a cutoff $\Lambda$ on $E_{\rm parton}$
  for an assumed integrated luminosity of $10\ifb$. Thin lines show a
  $\pm$10\% variation of $k_{\rm tr}$ about $M_D$, the straight line is
  $M_D^{\rm max}=\Lambda$. The leveling-off at $M_D^{\rm max}\approx \Lambda$
  reflects the gravitational UV fixed point.To enhance the reach we require
  $m_{\ell\ell}^{\rm min}{=}{\rm min}(M_D/3,2\tev)$.}\vskip-.4cm
\label{fig:discovery}
\end{figure*}

{\it Gravitational dilepton production.---} Now we turn to scenarios with
large extra dimensions where gravity propagates in the $D$ dimensional bulk
and matter fields live on a four-dimensional brane \cite{add}. The
four-dimensional Planck scale $M_{\rm Pl}$ is related to the size $L$ of the
$n$ extra dimensions and the $D$-dimensional Planck scale $M_D$ as $M_{\rm
  Pl}\sim M_D^2(M_DL)^n$. A fundamental Planck scale in the TeV range requires
$1/L\ll M_D$.  Virtual graviton effects, as opposed to real graviton emission,
are more sensitive to an underlying UV fixed point \cite{grw}.  Hence we
consider dilepton production through virtual gravitons.  To lowest order in
canonical dimension, this is generated through an effective dimension--8
operator in the effective action, involving four fermions and a
graviton~\cite{grw,gs,gps}.  Tree--level graviton exchange is described by an
amplitude ${\cal A} = {\cal S}\cdot {\cal T}$, where ${\cal T} = T_{\mu\nu}
T^{\mu\nu} - \frac{1}{2+n} T_\mu^\mu T_\nu^\nu$ is a function of the
energy-momentum tensor, and
\begin{equation} \label{S}
{\cal S}= \frac{S_{n-1}}{\mp^{2+n}} \; 
            \int_0^\infty d m \; m^{n-1}\, P(s,m)
\end{equation}
with $S_{n-1}=2 \pi^{n/2}/\Gamma(n/2)$ is a function of the scalar part
$P(s,m)$ of the graviton propagator~\cite{grw,gs,gps}. The integration over
the Kaluza-Klein (KK) masses $m$, which we can take as continuous, reflects
that gravity propagates in the higher-dimensional bulk. If the graviton
anomalous dimension is small $|\eta|\ll 1$, the propagator is
\begin{equation}\label{propIR}
P(s,m)=(s+m^2)^{-1}\,.
\end{equation}
This is valid if the relevant momentum transfer is $\ll M_D$. It is well
known that \eq{S} with \eq{propIR} is UV divergent for $n\ge 2$ \cite{grw}.
Implementing an UV cutoff $\Lambda$ \cite{gs} as the upper integration
boundary in \eq{S}, and using \eq{propIR}, gives
\begin{equation}
{\cal S}_\Lambda 
         = \frac{S_{n-1}}{n-2} \frac{1}{M^4_D} \; 
           \left( \frac{\Lambda}{M_D} \right)^{n-2} \;
\left[1+{\cal O}\left(\frac{s}{\Lambda^2}\right)\right]
\label{eq:s_theta}
\end{equation}
The strong cutoff dependence shows that \eq{S} with \eq{propIR} is dominated
in the UV by the Kaluza-Klein modes, if $n\ge 2$.  In Wilson's approach
\eq{ERG}, the graviton's anomalous dimension $\eta$ and its scale dependence
have to be taken into account.  Technically this is implemented by matching
the RG scale $k^2$ with $p^2=s+m^2$, the relevant scale of the propagator
entering \eq{S}, and $k_{\rm tr}$ of the order of the Planck scale.  For the
dressed propagator $1/(Z_G(k)\ p^2)$, this leads to a behaviour $\sim
p^{-2(1-\eta(p)/2)}$. With \eq{eta}, this becomes $\sim (p^2)^{-D/2}$ for
large $p^2$, as opposed to the $p^{-2}$ behaviour in the perturbative domain.
Applying this RG improvement to \eq{propIR}, we are lead to
\begin{equation}\label{propUV}
P(s,m)=\frac{k_{\rm tr}^{n+2}}{(s+m^2)^{n/2+2}}
\end{equation}
in the vicinity of an UV fixed point. The observation central for our purposes
is that \eq{S} with \eq{propUV} is finite even in the UV limit of the
integration.  The large anomalous dimension in asymptotically safe gravity
suffices to provide for a finite dilepton production rate.  \step

We emphasize that our RG improvement is conceptually different from the form
factor introduced in \cite{Hewett:2007st}. The latter employs the matching
$k^2 = s$, implying that Kaluza-Klein modes are treated
perturbatively. Consequently, and unlike here, the form factor method
\cite{Hewett:2007st} still requires an UV cutoff to render \eq{S} finite,
analogous to \eq{eq:s_theta} within effective field theory.  \step

{\it Results.---} The task to evaluate dilepton rates via the RG improvement
of \eq{S} is simplified by noticing that the anomalous dimension displays a
rapid cross-over from IR to UV scaling (see Fig.~\ref{Running}) in particular
in higher dimensions. Therefore we use \eq{propIR} in the IR regime where
$k<k_{\rm tr}$ and \eq{propUV} for the UV regime $k>k_{\rm tr}$, $k_{\rm
  tr}=M_D$. Since the integral at hadron colliders is sensitive to a
gluon--density induced bias towards the smallest possible $s$, we retain for
simplicity the leading term in $s/M_D^2$ in either regime.

In Fig.~\ref{fig:discovery} we display the discovery potential in $M_D$ at the
LHC. Taking into account the leading backgrounds we compute the minimal signal
cross section for which a $5\sigma$ excess is observed, assuming statistical
errors. (See \cite{gps} for technical details.)  This minimal cross section
translates into a maximum reach $M_D^{\rm max}$.  Consistency is checked by
introducing an artificial cutoff $\Lambda$ on the partonic energy \cite{gps},
setting the partonic signal cross section to zero for $E_{\rm
  parton}>\Lambda$.  This cutoff is not required in our approach and for
sufficiently large $\Lambda$, $M_D^{\rm max}$ becomes independent of it.  This
is nicely seen in Fig.~\ref{fig:discovery} once $\Lambda > M_D$. The fact that
$M_D^{\rm max}$ levels off at about $\Lambda \approx M_D^{\rm max}$ reflects
the onset of the underlying UV fixed point. To estimate uncertainties in our
approximations, we allow for a 10\% variation in $k_{\rm tr}$, leading to mild
variations in Fig.~\ref{fig:discovery} of a similar magnitude, slightly
increasing with $n$.

\begin{figure}[t]
\vskip-4mm
\includegraphics[width=.44\textwidth]{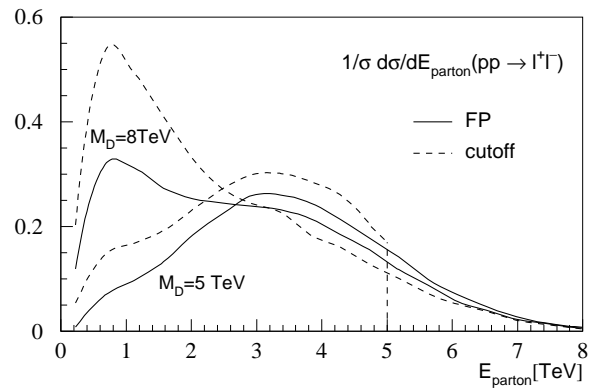}\vskip-4mm
\caption{Comparison of normalized distributions of the partonic energy 
  $E_{\rm parton}$ for the dimension--8 operator correction to Drell--Yan
  production at the LHC $(n=3)$. Full line: present work, dashed line:
  approximation \eq{eq:s_theta} with $\Lambda=M_D$.}
\label{fig:spectrum}\vskip-.6cm
\end{figure}

In Fig.~\ref{fig:spectrum} we show the normalized $E_{\rm parton}$
distributions for Drell--Yan production including KK gravitons for $n=3$ and
$M_D=5$ and $8$~TeV. In contrast to Fig.~\ref{fig:discovery} we do not apply
any $m_{\ell\ell}$ cut here. The solid curves represent our fixed point
analysis. The entire $E_{\rm parton}$ range contributes to the rate as long as
there is a sizeable parton luminosity. The dashed curves correspond to the
approximation \eq{eq:s_theta} with UV cutoff $\Lambda =M_D$.  Obviously, there
are no contributions above $E_{\rm parton}=\Lambda$.  The two sets of curves
do not scale in a simple manner because Standard Model and KK amplitudes
interfere.  For small $E_{\rm parton}$ the interference term is significant,
whereas for large $E_{\rm parton}$ there is hardly any SM contribution.\step

In Tab.~\ref{tab:rates}, we show the LHC production cross-section for
dileptons, $n=3$ and $6$. Our fixed point results are given in $a)$ and $b)$.
In $a)$, we retain the leading term of \eq{S} in $s/M_D^2$.  In $b)$, we
additionally implement the leading large-$s$ suppression starting at $M_D$. In
$c)$ we introduce a cutoff in $m$ and partonic energy at $M_D$, following
\cite{gps}.  For large $M_D\approx 5-8\tev$ the LHC has little sensitivity
to quantum gravitational effects and we find only small differences between
$a)$ and $b)$. Here, the difference between $b)$ and $c)$ is due to the
Kaluza Klein sector. For small $M_D\approx 2\tev$, $a)-c)$ lead to significant
differences which originate from physics at and beyond the fundamental Planck
scale, omitted in $c)$.\step

\begin{table}[t]
\begin{tabular}{|c|rrr|rrr|}  
   \hline
$ \sigma[{\rm fb}]$ & & $n=3$ &           
                       & & $n=6$ &                \\[1mm]
   $\mp$               & $2\tev$ & $5\tev$ & $8\tev$ 
                       & $2\tev$ & $5\tev$ & $8\tev$   \\[1mm]
  \hline
   $a)$           & 2270 & 1.41 & 0.0317 &  2220 & 1.36  & 0.031 \\
   $b)$ &  408 & 1.24 & 0.0317 &   398 & 1.21  & 0.031 \\
   $c)$           &  173 & 0.72 & 0.0204 &    66 & 0.28  & 0.008 \\
\hline
\end{tabular}  
\caption{Comparison of dilepton production rates at the LHC after 
acceptance cuts.  See main text for the definitions of $a)$, $b)$ and $c)$. 
All rates use  $m_{\ell\ell}^{\rm min}{=}{\rm min}(M_D/3,2\tev)$.}
\label{tab:rates}\vskip-.5cm
\end{table}

Finally, we comment on possible extensions. We have taken only the non-trivial
anomalous dimension of the graviton into account, which is the dominant effect
in asymptotically safe gravity. Vertex corrections can equally be studied
\cite{Percacci:2003jz}. Systematic expansions of Wilson's flow \eq{ERG} in
$N$-point functions work very well in other theories, $e.g.$ infrared QCD
\cite{Pawlowski:2003hq}.  For the physical observables studied here, we expect
vertex corrections to be subleading because the relevant momentum integrals
are dynamically suppressed above the Planck scale.  Our simple matching in the
cross-over regime can be extended by implementing the explicit RG running of
$\eta$, using the results of \cite{Litim:2003vp,Fischer:2006fz}.  The
variation in the matching scale (Fig.~\ref{fig:discovery}) should cover these
uncertainties.\step

{\it Conclusions.---} We have laid out a framework to study
quantum-gravitational effects at high energies within Wilson's renormalisation
group. We applied this to lepton pair production through virtual graviton
exchange in scenarios with large extra dimensions. The main new effects are
dictated by a gravitational UV fixed point above the Planck scale.
Remarkably, the renormalisation group improvement advocated here results in
finite cross sections for gravitational lepton pair production, and leads to
controlled experimental signatures at the LHC already at low integrated
luminosity.  This indicates that asymptotically safe gravity could be
detectable at hadron colliders, provided the fundamental scale of gravity is
as low as the electroweak scale.  

{\it Acknowledgements.---} The work of DFL is supported by an EPSRC Advanced
Fellowship.


\begin{thebibliography}{99}


\bibitem{add}
 N.~Arkani-Hamed, S.~Dimopoulos and G.~R.~Dvali,
  Phys.\ Lett.\ B {\bf 429}, 263 (1998);
 I.~Antoniadis, N.~Arkani-Hamed, S.~Dimopoulos and G.~R.~Dvali,
  Phys.\ Lett.\ B {\bf 436}, 257 (1998).

\bibitem{grw}
 G.~F.~Giudice, R.~Rattazzi and J.~D.~Wells,
  Nucl.\ Phys.\ B {\bf 544}, 3 (1999);
 T.~Han, J.~D.~Lykken and R.~J.~Zhang,
  Phys.\ Rev.\ D {\bf 59}, 105006 (1999).

\bibitem{real_kk}
 E.~A.~Mirabelli, M.~Perelstein and M.~E.~Peskin,
  Phys.\ Rev.\ Lett.\  {\bf 82}, 2236 (1999);
 T.~Han, D.~L.~Rainwater and D.~Zeppenfeld,
  Phys.\ Lett.\ B {\bf 463}, 93 (1999).
 L.~Vacavant and I.~Hinchliffe,
  hep-ex/0005033
 and
  J.\ Phys.\ G {\bf 27}, 1839 (2001);
 for an overview see \eg
 J.~Hewett and M.~Spiropulu,
  Ann.\ Rev.\ Nucl.\ Part.\ Sci.\  {\bf 52}, 397 (2002)

\bibitem{virtual_kk}
 J.~L.~Hewett,
  Phys.\ Rev.\ Lett.\  {\bf 82}, 4765 (1999);
 K.~M.~Cheung and G.~Landsberg,
  Phys.\ Rev.\ D {\bf 62}, 076003 (2000).


\bibitem{Weinberg79} S.~Weinberg, in {\it General Relativity: An
Einstein centenary survey}, Eds.~S.W.~Hawking and W.~Israel, Cambridge
University Press (1979), p.~790.

\bibitem{Reuter:1996cp}
  M.~Reuter,
  Phys.\ Rev.\  D {\bf 57}, 971 (1998).

\bibitem{Souma:1999at}
  W.~Souma,
  Prog.\ Theor.\ Phys.\  {\bf 102}, 181 (1999).

\bibitem{Lauscher:2001rz}
  O.~Lauscher and M.~Reuter,
  Class.\ Quant.\ Grav.\  {\bf 19}, 483 (2002).

\bibitem{Percacci:2003jz}
  R.~Percacci and D.~Perini,
  Phys.\ Rev.\  D {\bf 68}, 044018 (2003);
  A.~Codello and R.~Percacci,
  Phys.\ Rev.\ Lett.\  {\bf 97}, 221301 (2006);
  A.~Codello, R.~Percacci and C.~Rahmede,
  0705.1769 [hep-th].

\bibitem{Litim:2003vp}
  D.~F.~Litim,
  Phys.\ Rev.\ Lett.\  {\bf 92}, 201301 (2004);
  AIP Conf.\ Proc.\  {\bf 841}, 322 (2006).

\bibitem{Niedermaier:2006ns}
M.~Niedermaier,
  gr-qc/0610018;
M.~Niedermaier and M.~Reuter, Living Rev.~Relativity {\bf 9}, 5 (2006).


\bibitem{lattice}
  H.~W.~Hamber,
  0704.2895 [hep-th];
  J.~Ambjorn, J.~Jurkiewicz and R.~Loll,
  Phys.\ Rev.\ Lett.\  {\bf 93}, 131301 (2004).

\bibitem{Fischer:2006fz}
  P.~Fischer and D.~F.~Litim,
  Phys.\ Lett.\  B {\bf 638}, 497 (2006);
  AIP Conf.\ Proc.\  {\bf 861}, 336 (2006).

\bibitem{Wetterich:1992yh}
  C.~Wetterich,
  Phys.\ Lett.\  B {\bf 301}, 90 (1993).

\bibitem{Litim:2000ci}
D.~F.~Litim,
  Phys.\ Lett.\  B {\bf 486}, 92 (2000);
  Phys.\ Rev.\  D {\bf 64}, 105007 (2001);
  Nucl.\ Phys.\  B {\bf 631}, 128 (2002).


\bibitem{Freire:2000bq}
  F.~Freire, D.~F.~Litim and J.~M.~Pawlowski,
  Phys.\ Lett.\  B {\bf 495}, 256 (2000).

\bibitem{gs}
 G.~F.~Giudice and A.~Strumia,
  Nucl.\ Phys.\ B {\bf 663}, 377 (2003).

\bibitem{gps}
 G.~F.~Giudice, T.~Plehn and A.~Strumia,
  Nucl.\ Phys.\ B {\bf 706}, 455 (2005)



\bibitem{Hewett:2007st}
  J.~Hewett and T.~Rizzo,
  0707.3182 [hep-ph].

\bibitem{Pawlowski:2003hq}
  J.~M.~Pawlowski, D.~F.~Litim, S.~Nedelko and L.~von Smekal,
  Phys.\ Rev.\ Lett.\  {\bf 93}, 152002 (2004).

\end{thebibliography}
\end{document}